\documentclass[apjl]{emulateapj}
\usepackage{apjfonts}
\newcommand{\kms}{\,km\,s$^{-1}$}     
\newcommand{\ha}{\,H$\alpha$}     

\newcommand{\arcs}{$^{\prime\prime}$}
\newcommand{\arcm}{$^{\prime}$}

\journalinfo{To be published the Galaxy Evolution
Explorer (GALEX) Astrophysical Journal Letters Special Issue}
\shorttitle{GALEX observations of NGC~253 and M82}
\shortauthors{Hoopes et al.}

\begin{document}

\title{GALEX Observations of the Ultraviolet Halos of NGC~253 and M82}

\author{
Charles G. Hoopes\altaffilmark{1}, 
Timothy M. Heckman\altaffilmark{1}, 
David K. Strickland\altaffilmark{1}, 
Mark Seibert\altaffilmark{2}, 
Barry F. Madore\altaffilmark{3,4}, 
R. Michael Rich\altaffilmark{5}, 
Luciana Bianchi\altaffilmark{6}, 
Armando Gil de Paz\altaffilmark{3, 4}, 
Denis Burgarella\altaffilmark{7}, 
David A. Thilker\altaffilmark{1}
Peter G. Friedman\altaffilmark{2},
Tom A. Barlow\altaffilmark{2},
Yong-Ik Byun\altaffilmark{8}, 
Jose Donas\altaffilmark{7},
Karl Forster\altaffilmark{2}, 
Patrick N. Jelinsky\altaffilmark{9},
Young-Wook  Lee\altaffilmark{8},
Roger F. Malina\altaffilmark{7},
D. Christopher Martin\altaffilmark{2}, 
Bruno Milliard\altaffilmark{7},
Patrick F. Morrissey\altaffilmark{2}, 
Susan G. Neff\altaffilmark{10},
David Schiminovich\altaffilmark{2},
Oswald H. W. Siegmund\altaffilmark{9}, 
Todd Small\altaffilmark{2},
Alex S. Szalay\altaffilmark{1}, 
Barry Y. Welsh\altaffilmark{9},
and Ted K. Wyder\altaffilmark{2}
}

\altaffiltext{1}{Department of Physics and Astronomy, The Johns Hopkins
University, Homewood Campus, Baltimore, MD 21218}

\altaffiltext{2}{California Institute of Technology, MC 405-47, 1200 East
California Boulevard, Pasadena, CA 91125}

\altaffiltext{3}{Observatories of the Carnegie Institution of Washington,
813 Santa Barbara St., Pasadena, CA 91101}

\altaffiltext{4}{NASA/IPAC Extragalactic Database, California Institute
of Technology, MC 100-22, 770 S. Wilson Ave., Pasadena, CA 91125}

\altaffiltext{5}{Department of Physics and Astronomy, University of
California, Los Angeles, CA 90095}

\altaffiltext{6}{Center for Astrophysical Sciences, The Johns Hopkins
University, 3400 N. Charles St., Baltimore, MD 21218}

\altaffiltext{7}{Laboratoire d'Astrophysique de Marseille, BP 8, Traverse
du Siphon, 13376 Marseille Cedex 12, France}

\altaffiltext{8}{Center for Space Astrophysics, Yonsei University, Seoul
120-749, Korea}

\altaffiltext{9}{Space Sciences Laboratory, University of California at
Berkeley, 601 Campbell Hall, Berkeley, CA 94720}

\altaffiltext{10}{Laboratory for Astronomy and Solar Physics, NASA Goddard
Space Flight Center, Greenbelt, MD 20771}

\begin{abstract}

We present Galaxy Evolution Explorer (GALEX) images of the
prototypical edge-on starburst galaxies M82 and NGC~253. Our initial
analysis is restricted to the complex of ultraviolet (UV) filaments in
the starburst-driven outflows in the galaxy halos. The UV luminosities
in the halo are too high to be provided by continuum and line emission
from shock-heated or photoionized gas except perhaps in the brightest
filaments in M82, suggesting that most of the UV light is the stellar
continuum of the starburst scattered into our line of sight by dust in
the outflow. This interpretation agrees with previous results from
optical imaging polarimetry in M82. The observed luminosity of the
halo UV light is $\la0.1\%$ of the bolometric luminosity of the
starburst. The morphology of the UV filaments in both galaxies shows a
high degree of spatial correlation with \ha\ and X-ray emission. This
indicates that these outflows contain cold gas and dust, some of which
may be vented into the intergalactic medium (IGM). UV light is seen in
the ``\ha\ cap'' 11 kpc North of M82. If this cap is a result of the
wind fluid running into a pre-existing gas cloud, the gas cloud
contains dust and is not primordial in nature but was probably
stripped from M82 or M81.  If starburst winds efficiently expel dust
into the IGM, this could have significant consequences for the
observation of cosmologically distant objects.

\end{abstract}


\section{Introduction}

Many local starburst galaxies have galactic-scale outflows of
metal-enriched gas, called starburst superwinds, which are
driven by the stellar winds and supernovae of numerous massive
stars ({\it e.g.,} Heckman, Armus, \& Miley 1990). These outflows
contain a hot ($10^7$~K), metal-enriched wind fluid as well as
entrained cooler gas and dust \citep{ss00}. It is likely that the
hot gas can escape from the potential well of the parent galaxy,
enriching the intergalactic medium (IGM) with metals and energy
\citep{hlsa00}. It is not clear whether the colder gas can escape, and
this is an important question since it would mean that superwinds
also enrich the IGM with dust, which could affect observations
of high-redshift objects \citep{a99,adb99,a01,hlsa00}. This question
is even more crucial since similar outflows are now known to be common
in high-redshift starbursts \citep{pet01,sspa03}.

Several lines of evidence suggest that superwinds contain dust
\citep{sb98,hlsa00}.  Optical imaging polarimetry shows light
scattered by dust in the halos of starburst galaxies, including M82
(e.g Scarrott, Eaton, \& Axon 1991; Alton et al. 1994). Far-IR and
sub-mm imaging reveal thermal emission from extraplanar dust in
several edge-on starburst galaxies (Alton, Davies, \& Bianchi
1999). Finally, the strong correlation between the strength of the
blueshifted interstellar Na~D line and the line-of-sight reddening in
superwinds \citep{hlsa00}, strongly suggests the dust is actually
outflowing. What remains unclear is the physical relationship between
the cool dust-bearing gas and the warm and hot gas probed by optical
lines and X-rays respectively. Comparison of sensitive, high
resolution images of the dusty material with \ha\ and X-ray emission
images would shed light on this relationship.  Dust is highly
reflective in the ultraviolet (UV) \citep{d03}, so imaging of
starburst superwinds in the UV can trace the location of dust, if one
can account for UV emission by photoionized or shock-heated
gas. Indeed, Ultraviolet Imaging Telescope (UIT) near-UV data for M82
show evidence of UV light in the halo corresponding to known \ha\
features \citep{marcum01}.  Here we present Galaxy Evolution
Explorer (GALEX) ultraviolet images of two prototypical starburst
superwind galaxies: NGC~253 and M82. The images reveal prominent UV
light in the superwind region. Our goal is to understand the origin of
this light.

\section{Observations}

\begin{figure*}
\plottwo{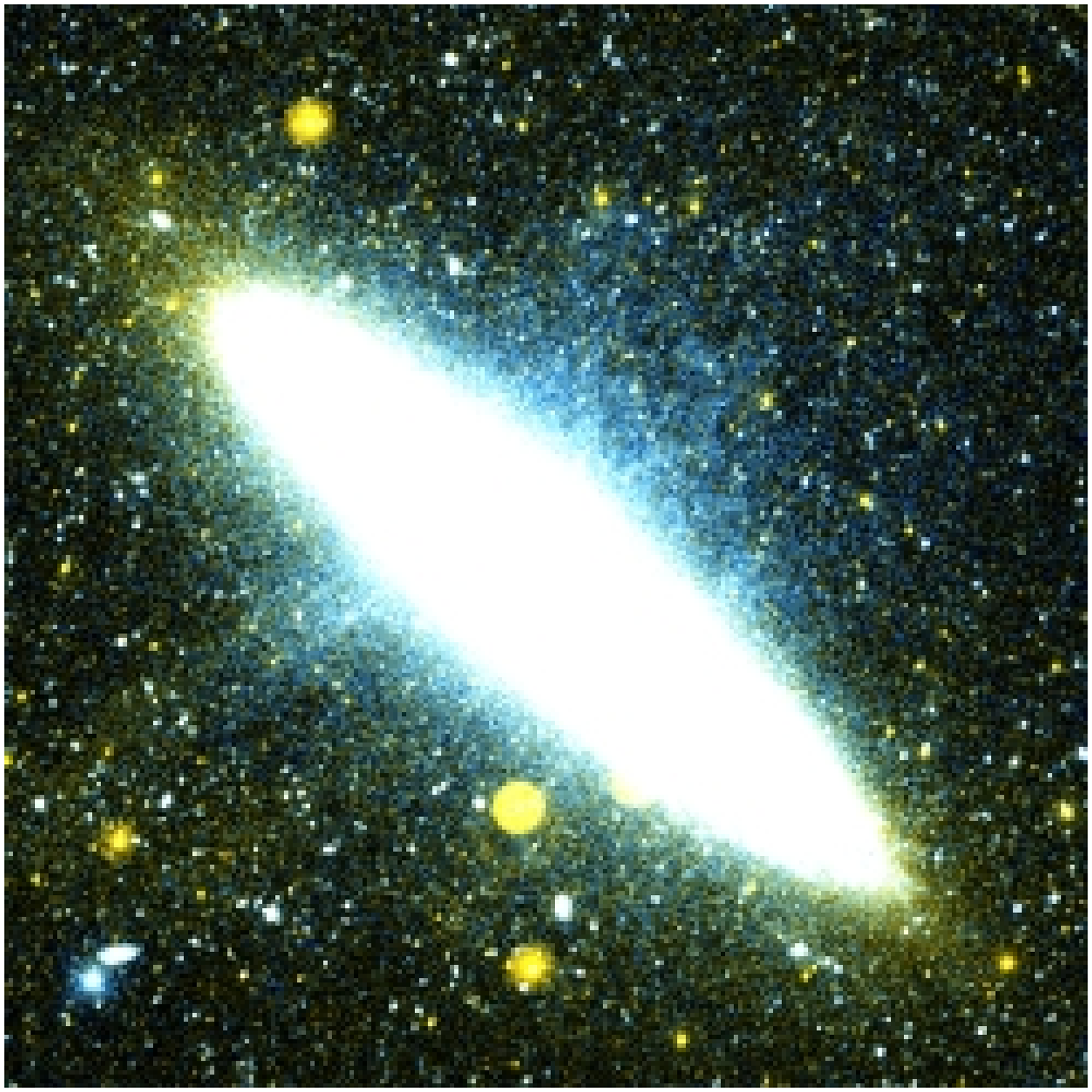}{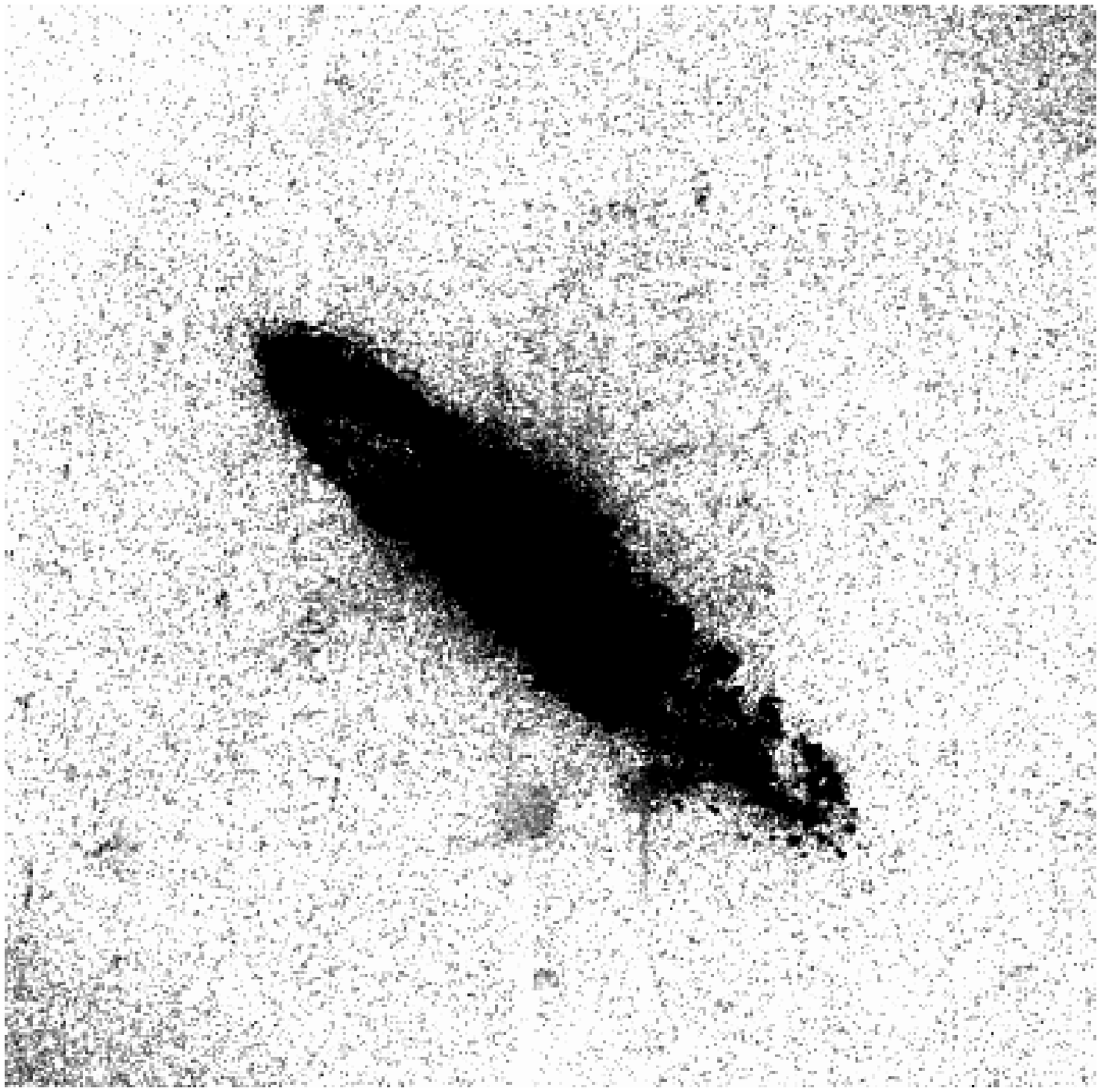}
\caption{
NGC~253 in UV and \ha. The left panel shows a two-color image, with
GALEX NUV in red and FUV in blue. The right panel shows the
continuum-subtracted \ha\ image. The images are $30^{\prime}$ (22.7
kpc) on each side, with North up and East on the left, and are aligned
with each other. The intensity scales in both panels are
logarithmically scaled and stretched to emphasize the faint, diffuse
emission, so the bright disk of the galaxy is saturated.  }
\label{n253fig}
\end{figure*}

NGC~253 was observed by GALEX on 2003 October 13 for 3289 seconds.
M82 was observed by GALEX on 2003 December 8 for 3083 seconds.  The
GALEX data include far-ultraviolet (FUV; $\lambda_{eff}=1528$~\AA,
$\Delta\lambda=268$~\AA) and near-ultraviolet (NUV;
$\lambda_{eff}=2271$~\AA, $\Delta\lambda=732$~\AA) images with a circular
field of view with radius $\sim38$\arcm. The spatial resolution is
$\sim5$~\arcs. Details of the GALEX instrument and data
characteristics can be found in \citet{dcm04} and
\citet{pm04}.

We also use previously obtained H$\alpha$ data. The \ha\ image of
NGC~253 is described in detail in \citet{hwg96}. The \ha\ image of M82
is part of a mosaic of the M81-M82 system taken with the Burrell-Schmidt
telescope at KPNO, and is described in \citet{gwth98}.

\section{Analysis}

\subsection{Ultraviolet Morphology}

Figure~\ref{n253fig} compares the two-color GALEX image of NGC~253
with the \ha\ image. Extended \ha\ emission was noted by
\citet{sh02}. Diffuse emission extending several kpc into the halo on
both sides of the disk, north-east of the galaxy center with the brightest and
more extended emission toward the east end of the disk. \citet{sh02}
found that the X-ray emission matched the \ha\ emission in
morphology. These features are also visible in the GALEX images.

\begin{figure*}
\plottwo{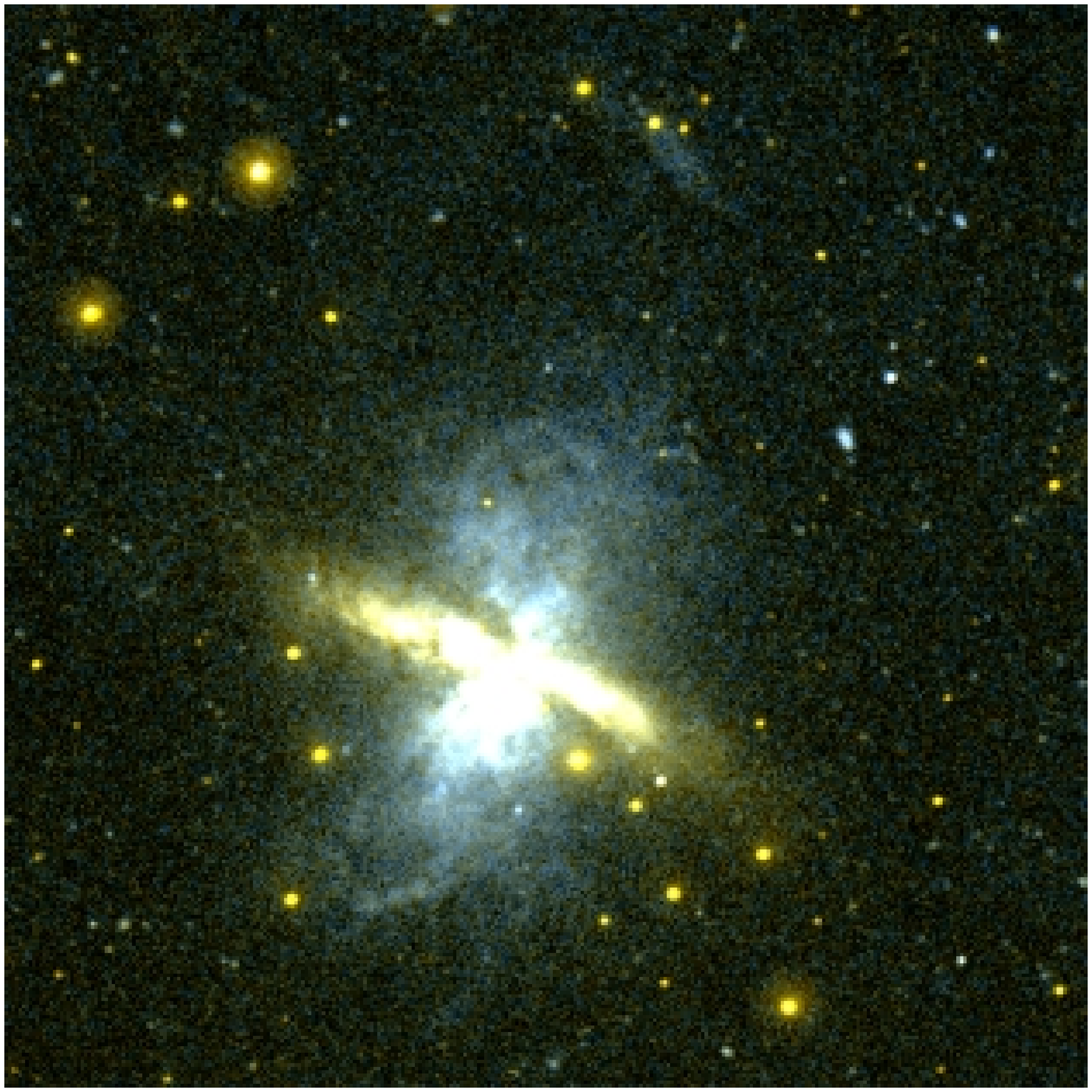}{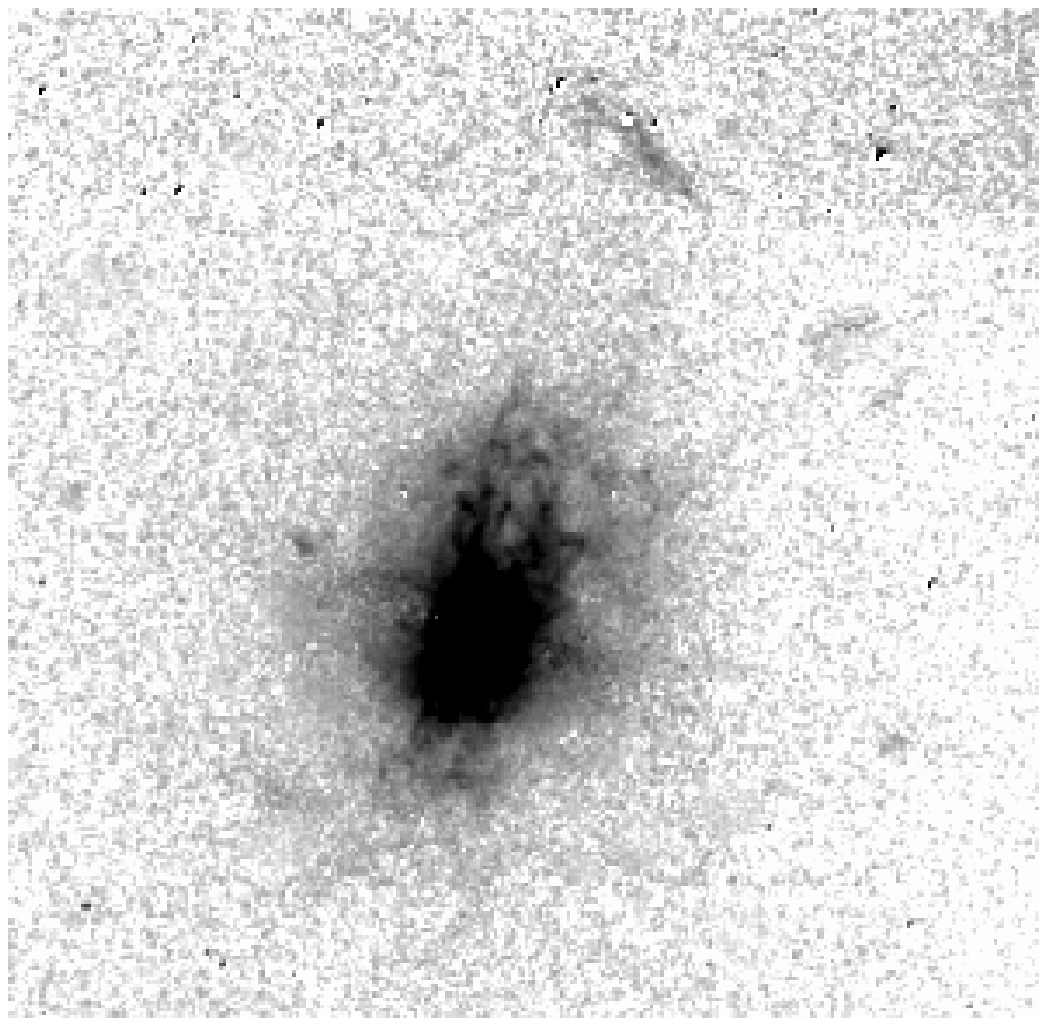}\\
\caption{
M82 in UV and \ha. The left panel shows a two-color image, with GALEX
NUV in red and FUV in blue. The right panel shows the
continuum-subtracted \ha\ image. The images are $21^{\prime}$ (22.0
kpc) on each side, with North up and East on the left, and are aligned
with each other. The intensity scales in both panels are
logarithmically scaled and stretched to emphasize the faint, diffuse
emission, so the bright disk of the galaxy is saturated.  }
\label{m82fig}
\end{figure*}

Figure~\ref{m82fig} shows the UV and \ha\ images of M82 (in the same
manner as Figure~\ref{n253fig}). The M82 images show a bright, complex
network of filaments, very different in appearance from NGC~253. The
morphology in the UV and \ha\ images is strikingly similar. Prominent
\ha\ filaments are seen perpendicular to the disk on both sides,
surrounded by a lower surface brightness component of diffuse light
(see also Ohyama et al. 2002). The filaments are also visible in the
UV, but there is less contrast between the filaments and the diffuse
UV light. The GALEX images are much more sensitive than the earlier UIT NUV
image \citep{marcum01}, and show that the UV-\ha\ correlation in
morphology extends to very faint \ha\ filaments, and exists in the FUV
as well (which was not detected by UIT).  \citet{sh04} noted
that the X-ray and \ha\ morphology are similar on all scales, and this
is also true for the UV light. The \ha\ and X-ray ``cap''
\citep{db99,lhw99} 11~kpc above the North side of the disk is visible
in both GALEX bands. We will address the UV properties of these and
other nearby starbursts, including possible reasons for the striking
differences between M82 and NGC~253, in a forthcoming paper (Hoopes et
al., in preparation).

\subsection{Luminosities and Flux Ratios}

\begin{deluxetable*}{lccccccccc}
\tabletypesize{\small}
\tablecaption{Measured Properties\tablenotemark{a} \label{lum}}
\tablewidth{0pt}
\tablehead{
\colhead{Galaxy} & \colhead{Distance} & 
\colhead{$L_{H\alpha}$ Halo} & \colhead{$L_{H\alpha}$ Total} & 
\colhead{$L_{NUV}$ Halo} & \colhead{$L_{NUV}$ Total} & 
\colhead{$L_{FUV}$ Halo} & \colhead{$L_{FUV}$ Total} & 
\colhead{$L_{Bol}$\tablenotemark{b}} & \colhead{$\beta$\tablenotemark{c}} \\
\colhead{} & \colhead{(Mpc)} & 
\colhead{(erg~s$^{-1}$)}& \colhead{(erg~s$^{-1}$)}& 
\colhead{(erg~s$^{-1}$)} & \colhead{(erg~s$^{-1}$)} & 
\colhead{(erg~s$^{-1}$)} & \colhead{(erg~s$^{-1}$)} & 
\colhead{(erg~s$^{-1}$)} &  \colhead{}
}
\startdata
NGC~253 & 2.6 & $1.5\times10^{39}$ & $3.8\times10^{40}$ & $3.1\times10^{40}$ & $4.5\times10^{41}$ & $2.1\times10^{40}$ & $2.2\times10^{41}$ & $7.8\times10^{43}$ & -1.5\\
M82     & 3.6 & $1.3\times10^{40}$ & $6.1\times10^{40}$ & $1.5\times10^{41}$ & $3.5\times10^{41}$ & $7.1\times10^{40}$ & $1.1\times10^{41}$ & $2.0\times10^{44}$ & -0.6\\
\enddata
\tablenotetext{a}{The measured luminosities were corrected for foreground Galactic
extinction. Calibration uncertainties are $\sim10$~\% in the UV bands
\citep{pm04} and are of similar magnitude in H$\alpha$.}
\tablenotetext{b}{The bolometric luminosities were taken from \cite{sh00} (NGC~253) and
\citet{mrrk93} (M82).}
\tablenotetext{c}{The spectral slope, defined via
$F_{\lambda}\propto\lambda^{\beta}$, was estimated from the FUV/NUV flux
ratio following \citet{k04}.}
\end{deluxetable*}

Table~\ref{lum} compares the UV and \ha\ luminosities of the halo with
the total and bolometric luminosities. The measurements have been
corrected for Galactic foreground extinction using $E(B-V)=0.019$ for
NGC~253 and $E(B-V)=0.493$ for M82 \citep{schlegel98}. A correction factor
of 0.59 has been applied to remove [\ion{N}{2}] from the \ha\
flux. The extraplanar UV light in both cases is less than $0.1\%$ of
the bolometric luminosity of the starburst. The observed halo
luminosity is $7\%$ ($10\%$) of the total {\it observed} NUV(FUV)
luminosity of NGC~253, and for M82 it is $43\%$ ($65\%$).  The \ha\
luminosity of the halo is $4\%$ of the total \ha\ luminosity for
NGC~253 and $21\%$ for M82.

\begin{figure*}
\epsscale{1.12}
\plotone{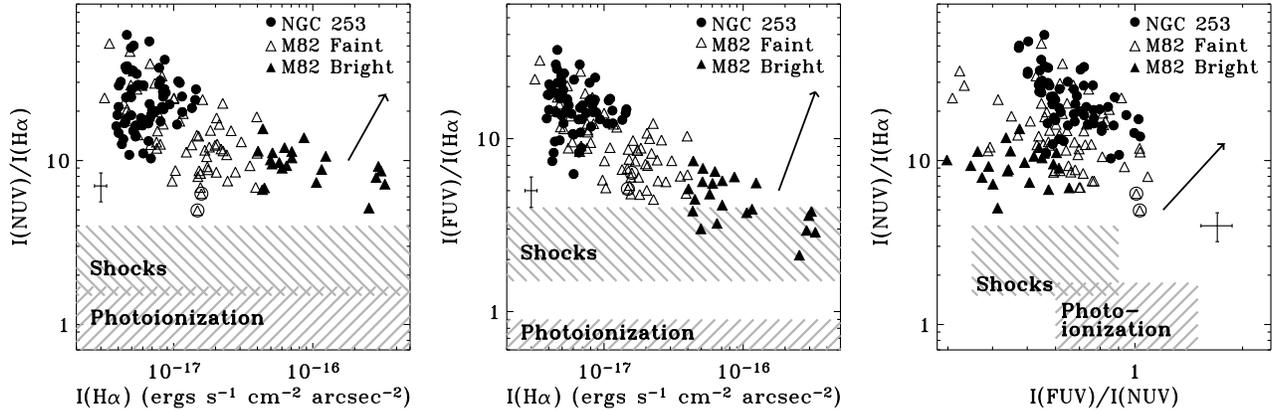}
\caption{
Comparison of measured ratios with model predictions. The points are
the measured values, and have been corrected for Galactic foreground
extinction using the extinction law of \citet{ccm89}.  The hatched
regions indicate the range of model predictions. The models are
described in the text. We have not attempted to correct for internal
extinction. Reddening vectors for $E(B-V)=0.21$ (measured in the M82
wind by Heckman et al. 1990) are shown in each panel, assuming the
\citet{c01} starburst extinction law. The M82 points are separated into
bright and faint based on their \ha\ surface brightness, with the
division occurring at
$I(H\alpha)=4\times10^{-17}$~ergs~s$^{-1}$~cm$^{-2}$~arcsec$^{-2}$. The
two circled triangles are in the M82 \ha\ cap. Representative error bars are shown in each panel.}
\label{ratios}
\end{figure*}

Figure~\ref{ratios} shows flux ratios measured in square regions
30\arcs\ on each side. The GALEX monochromatic fluxes were multiplied
by the effective filter bandpass to give units of
ergs~cm$^{-2}$~s$^{-1}$.  Figure~\ref{ratios} also shows model
predictions for the continuum (Balmer, Bremsstrahlung, and two-photon)
and line emission of shock-heated and photoionized gas \citep{ds96,
f96}. The shock models span shock velocities from 100 to 900 \kms, and
include both shock and precursor emission. The photoionization models
are of spherically symmetric clouds ionized by a central source (the
UV continuum of the ionizing source is not included in the model
predictions), and span stellar temperatures ranging from 30000~K to
50000~K and electron densities from 0.1 to 10~cm$^{-3}$. Solar
abundances were assumed in both cases.

Most of the regions have too much UV light (relative to \ha) to be
explained by nebular emission alone. The observed FUV/\ha\
ratios in some of the brighter regions in the M82 halo are consistent
with a significant contribution from shock ionization, but these
values have {\it not} been corrected for extinction intrinsic to the
wind. The optical spectrum of the M82 wind indicates a reddening of
$E(B-V)\ga0.21$~mag \citep{ham90}, which would increase the observed
FUV/\ha\ ratios by a factor of $\ga3.9$ ($\ga2.6$ for NUV/\ha). This
implies that the wind is substantially brighter in FUV and NUV than
would be possible for photoionized or shock heated gas. The absence of
\ion{O}{6} emission seen in {\em Far Ultraviolet Spectroscopic
Explorer} data limits shock speeds in the bright M82 filaments to
$v_s\le 160$~km~s$^{-1}$ \citep{hhsh03}, much slower than the wind
velocity ($v\ge10^3$~km~s$^{-1}$, Strickland \& Stevens 2000). Taken
together, these facts imply that another source is required to explain
the excess extraplanar UV light. The diffuse morphology argues against
star formation in the wind as the source. The most likely remaining
mechanism is scattering of stellar continuum from the starburst by
dust in the halo.

\section{Discussion}

The spectral slopes $\beta$ in the halo implied by the observed
FUV/NUV ratio are listed in Table~\ref{lum}. The values are redder
than an unreddened starburst ($-2.5\le\beta\le-2.0$), and in fact
agree well with observed ({\it i. e.,} reddened) values of local
starbursts ($-2.0\le\beta\le-0.6$; Meurer, Heckman, \& Calzetti
1999). While it is clear that dust in a starburst environment may have
properties that differ from the standard Galactic dust models
\citep{gcw97,p00}, the FUV/NUV ratio is in general agreement with the
dust scattering models of \citep{d03} in which the dust albedo is
greater in the NUV than in the FUV.

In a sample of local star forming galaxies, \citet{b02} found that the
observed \ha\ flux was on average 2.5-5.0\% of the observed UV flux
near 2000~\AA\ (similar to the GALEX NUV band). The observed \ha\
flux of the NGC~253 halo is 5\% of the NUV flux, and for M82 the
corresponding value is 12\%.  This may indicate that not all of
the \ha\ emission in the M82 halo is scattered light, although
polarization measurements indicate the presence of some scattered
light in the \ha\ filaments and the more extended diffuse \ha\
component of M82 ({\it e. g.,} Scarrott et al. 1991). Scattered light
could be a substantial component of the fainter NGC~253 halo.

\cite{lhw99} suggested that the \ha\ cap near M82 is the result of a
collision between the hot wind fluid and a pre-existing neutral
cloud. If this scenario is correct, our results imply that the cloud
is not primordial since it contains dust. The cloud may have been
pushed out of M82 by the starburst wind or stripped from either M82 or
M81 by the tidal interaction between the two galaxies.

Our results establish for the first time a close morphological
correspondence between the dust and the hotter phases of the winds
probed in \ha\ and X-ray emission. We have direct evidence that the
hotter gas is outflowing ({\it e.g.,} Strickland \& Stevens 2000), so
the new UV images provide further evidence for outflowing dust
\citep{hlsa00}.  This is consistent with the idea that the cool dusty
material is ambient interstellar gas in the disk or inner halo that
has been has been entrained and accelerated by the hot outflowing gas
generated in the starburst.

If this dust is ejected into the IGM, there could be important
implications for cosmological observations. While the dust density is
small, over cosmological distances the resulting extinction could be
significant \citep{a99,hlsa00}.  \citet{abd01} point out that
intergalactic dust may affect the determination of the evolution of
the cosmic star formation rate, for example. More work is needed to
understand the effects of intergalactic dust.

\acknowledgments

We appreciate the helpful comments from the referee, Giuseppe Gavazzi.
We thank Daniela Calzetti, Cristina Popescu, and Richard Tuffs for
useful suggestions, and Ren\'e Walterbos and Bruce Greenawalt for
their part in obtaining and reducing the \ha\ data. GALEX (Galaxy
Evolution Explorer) is a NASA Small Explorer, launched in April
2003. We gratefully acknowledge NASA's support for construction,
operation, and science analysis for the GALEX mission, developed in
cooperation with the Centre National d'Etudes Spatiales of France and
the Korean Ministry of Science and Technology. The grating, window,
and aspheric corrector were supplied by France. We acknowledge the
dedicated team of engineers, technicians, and administrative staff
from JPL/Caltech, Orbital Sciences Corporation, University of
California, Berkeley, Laboratory Astrophysique Marseille, and the
other institutions who made this mission possible.


\end{document}